\documentstyle[aps,preprint]{revtex}

\begin{document}
\title{Metallic sulphur. ``Electronic'' mechanism of superconductivity?}
\author{V.N.Bogomolov}
\address{Russian Academy of Sciences\\
A.F. Ioffe Physicotechnical Institute\\
St. Petersburg, Russia}
\maketitle

\begin{abstract}
{It is shown that the rapid increase of the superconducting transition
temperature $T_c$ of sulphur with increasing pressure above 93 GPa does not
contradict with some hypothetical ``electronic'' mechanism of
superconductivity with participation of the electron interaction energy
fluctuations. Such ``electronic'' mechanism is supposed to be intrinsic
property of the molecular condensates and corresponds to very high $T_c$.
The low $T_c$ of sulphur (10 -17)K is likely connected with the magnetic
properties of the sulphur atoms and molecules. The equation of state for
sulphur is obtained. The molar volume of sulphur at metallization is 10 cm}$%
^3${/mol. The principal difference between the ''physical'' and the
''chemical'' type bonds are discussed. Under some pressure one bond type is
changed by another and }${T}_c${\ may have an extremum (transition from the
Bose condensation to the BCS superconductivity).}
\end{abstract}

In \cite{1,2} assumption has been made that the optical properties of
metallic xenon \cite{3} is a manifestation of the superHTSC. In \cite{4} the
superconductivity of sulphur has been found experimentally . It appears
simultaneously with the metallization at $93\,GPa$ ( $T_c\sim ~(10\div 17)K$
at the pressure $P~\sim (93\div 157)GPa$ ). Nevertheless low $T_c$ does not
exclude the existence of the ``electronic'' mechanism of superconductivity
if one take into account: 1) magnetic properties of molecules or atoms (m/a)
of sulphur, and 2) peculiarities of the $T_c(P)$ dependencies: a)
simultaneous appearance of metallization and superconductivity, b) increase
of $T_c$ with increasing pressure, c) fast increase of $T_c$ $(13\div 17)K$
with pressure above $155\,GPa\,(155\div 157)GPa.$

Molecular condensates (MC) with the ''physical'' type bond has a principal
difference from covalent substances (the ``chemical'' type bond ). In some
cases the description of the MC as a weak coupled systems with well defined
localized states (uncorrelated electron systems of m/a) seems to be more
adequate than the band theory. It may reveal a new properties of such
materials.

The most significant property of the MC is a closeness of the interatomic
distances $2r_2$ to the diameter of the excited state orbital of m/a $%
2r_2=e^2/(E_2-E_1)$ , $E1=e^2/2r_1$ - the ionization potential of m/a, $E_2$
- the energy difference between the ground and the excited states of m/a for
the MC. It is important that $r_2/r_1=y^{1/3}\gg 1$ - the main parameter of
the MC. The largest fraction of m/a of the MC are in the ground state, and
their electronic systems are not correlated. Interaction energy of these
electronic systems is about $w=e^2/2(r_2-r_1)$ and has a fluctuational
nature (it is a result of an ``accidental correlations''). It is the main
interaction which determines all properties of the MC. It determines the
population $X$ of the excited state orbitals and the mean bond energy,
metal-dielectric interaction, metallization, superconductivity \cite{1,2}.

Assuming that the transition to the excited state of m/a of a condensate
occurs as a result of perturbation $w$ of a fluctuational nature, we
calculate the population $X$ of the excited state orbitals according to the
formula:
\[
X=0,25\exp \left( -\frac{E_2}w\right) ,\;w=\frac{e^2}{2\left( r_2-r_1\right)
},\;X\sim \exp \left( -\frac{r_2}{r_1}\right) .
\]

It should be noted that $X$ well corresponds to the probability to find an
electron of the stable excited state at the point $r_2$. Such situation is
characteristic of the MC with the interatomic distances ~$\sim 2r_2$ only.
It is a ``physical'' type bond (in contrast to the ''chemical'' type bond,
which is realized via the correlated electron systems of m/a and with the
electronic properties governed by the band structure).

>From the discussed point of view $NX$ m/a are in the excited
(``metallized'') state and $N(1-X)$ m/a are in the ground state in MC at
every moment. It is very important that the interaction energy of electrons $%
w$ is a result of the interatomic pair interaction. The mean bond energy of
MC is determined by the bond energy of $NX/2$ virtual molecules of excimers
(m/a)$_2$ bonded by pair of electrons at excited states (the Frenkel
biexcitons). It leads to a possibility of the ``electronic'' type
superconductivity simultaneously with metallization. For example condensed
hydrogen $H_2$ is a typical MC and the bond energy and possible
superconductivity must be determined by $0.5NX(H_2)_2$ virtual molecules.

The main parameter of such a weak coupled MC is high $w$, but not low $kT$
as for strongly coupled covalent substances, and it is the compression, not
the thermal energy, which is a principal external parameter. $X$ is
proportional to the compression and at $X=X_p$ --- the percolation threshold
($\sim 0.12$) --- the metallization occurs (as the first stage of the Mott
transition at $y~\sim 12\div 15$ ). In the ``gas'' of molecular type virtual
pairs of electrons the Bose condensation, simultaneously with the
metallization at corresponding $T_B\sim ~X^{2/3}\sim {\rm ~constant}$ for
various MC, takes place. The next step is to substitute the dynamical
properties of lattice by that of the electronic system and $kT$ by $w$ for
description of the electromagnetic properties of the MC. So, the
superconducting transition temperature (by analogy the BCS formula) is:

\[
Tc~\sim 0.5w_mX=0.92E_1(y^{1/3}-1)^{-1}\exp \left[ -\left(
y^{1/3}+y^{-1/3}\right) \right]
\]
and $T_c$ is an increasing function of pressure. Magnetic susceptibility of
the monatomic MC is $\chi \sim ~1/w\ll 1/kT$.

The final stage of the Mott transition under increasing pressure ($y<12.2$)
is a transformation of the ``disordered'' MC into the ordinary metal with
well defined band structure (without any percolation processes). Molecular
type pairs are ``dissolved''. It is a transition of the ``physical'' type
substance into the ``chemical'' one under pressure and the BCS description
of the ordinary metals is valid ( ``phononic'' mechanism of
superconductivity , LTSC). The conclusion is that the pressure dependence of
$T_c$ must have a maximum. Such evolution is supposed to be for hydrogen
under pressure: transformation of diatomic MC into monatomic metal. It is
likely that the HTSC materials are in some intermediate state of the Mott
transition ( the electron concentration is about $XN\sim 0.12N$ and the
oxygen ion radius is close to that of the excited orbital ).

An attempt to construct some ``electronic'' mechanism (``electron pairing
induced by correlated charge fluctuations'') within band approximation is
made in \cite{5}. It seems more appropriate to search an ``electronic''
mechanism for MC with ''disordered'' localized electron excitations and well
defined ``molecules''. The Bose condensation is more probable for the
disordered systems than for the regular lattices.

The experimental data on $T_c$ of sulphur under pressure \cite{4} correspond
to the hypothetical picture presented above. Fast increase of $T_c$ with
increasing pressure above $155\,GPa$, found in the experiment \cite{4}, was
interrupted by the limited possibilities of the apparatus. Nevertheless
these results may be considered as a manifestation of the ``electronic''
mechanism if one takes into account the magnetic properties of m/a of
sulphur. Sulphur is a typical monatomic MC: $E_1=10.36\,eV$ ; $E_2=6.52\,eV$
; $r_1=0.692\,\AA $; $r_2=1.87\,\AA $ ($4s$ orbital ). The latter is close
to $1.85\AA $ -- the mean radius of atoms in the condensate; $y_0=19.3$. The
contribution to the magnetic susceptibility of $N(1-X)$ sulphur atoms in the
ground state is ~$\sim 1/w\ll 1/kT$, so that the macroscopic paramagnetism
may be neglected. Superconductivity appears simultaneously with
metallization, and $T_c$ increases with the increasing pressure.

The equation of state (EOS) for sulphur may be obtained from the EOS for
xenon \cite{2}:
\[
P_c(y)=P_0E_1^4\int (y-1)^{-7/3}\exp \left\{ {-2}\left[ {(\ y-1\ )}^{{1/3}}{%
\ +\ (\ y-1\ )}^{{-1/3}}\right] \right\} {\ }dy=P_1(E_1)F(y)+P_2
\]
with $P_1=3.7578024\,10^6(GPa)$ ; $P_2=-1.294312(GPa)$ ; $E_1=12.127\,eV$.

According to \cite{4}, the metallization of sulphur occurs at $P=93\,GPa$
and at $y_m=12.2$ \cite{2} (for close packed latices). At the normal
conditions $P=10^{-4}GPa$, and the $y$ parameter value is $19.3$. For this
case EOS parameters are: $P_1=1.8086688\,10^6(GPa)$ and $P_2=-29.0228679(GPa)
$. It is important that $P_{1Xe}/P_{1S}=2.07$, which is close to $%
(E_{1Xe}/E_{1S})^4=1.88$. The molar volume at the metallization, $%
V_0=16(12.2/19.3)~10cm^3/mol$, is close to the molar refractivity (the
Herzfeld criterion). Sulphur is a typical MC and may be described with the
approach discussed in \cite{1.2}.

As a next verification of the model, one can compare pressure dependencies
of $T_{c\,\exp }$ \cite{4} and $T_{c\,cal}$. It is clear, that we should
introduce low prefactor $Z$ in the formula for $T_c$ to account for some
microscopic magnetic properties of sulphur atoms (``paramagnetic
impurities'').

\begin{tabular}{llllllll}
$y$ & $19.3$ & $12.2$ & $11.86$ & $10.50$ & $11.08$ & $10.89$ & $10.57$ \\
$P\,(GPa)$ & $1\,10^{-4}$ & $93$ & $104$ & $120$ & $134$ & $142$ & $155$ \\
$T_{ce}(K)$ & $-$ & $10.1$ & $10.8$ & $11.8$ & $11.9$ & $12.3$ & $14.0$ \\
$T_{cc}(K)$ & $-$ & $10.4$ & $10.8$ & $12.4$ & $11.7$ & $11.9$ & $12.4$%
\end{tabular}

There is a satisfactory agreement between the $T_{cc}$ and $T_{ce}$ values
(at $Z\sim 0.002$) up to $142\,GPa$. At $P>155\,GPa$ the phase transition of
the sulphur lattice occurs and a significant rise of $T_c$ was discovered ($%
17\,K$ at $157\,GPa$). The huge discrepancy in the values of $T_{cc}$ and $%
T_{ce}$ ($4800\,K$ and $10\,K$) is not a catastrophe. In the sulphur, MC
magnetic virtual excimer molecules $(S)_2$ (interatomic distances $3.7\,\AA $
against $1.9\,\AA $ for $S_2$) exist according the model discussed. The size
of such magnetic molecules is about the coherent length for a superconductor
with $T_c\sim ~5000K$. The paramagnetic limit for superconductors is $%
H_p=1.8\,10^4T_c$ Oersted. This value is about the atomic magnetic fields
for $T_c\sim 5000K$ (the same situation is for the $(Xe)_2$ molecules in the
$Xe$-condensate, for the $(H_2)_2$ molecules in the $H_2$ -condensate, for
the $(O_2)_2$ molecules in the liquid oxygen). Moreover, $N(1-X)$ sulphur
atoms in the ground state are some kind of paramagnetic impurities too. It
is known to lower $T_c$ very much \cite{6}. Weak paramagnetism ($\sim 1/w$)
may be changed by some collective magnetic effects, if to increase the
interatomic interaction, and to substitute the mean fluctuation energy $w$
by the constant energy $w_0$ (transition to the ``chemical'' type bond). In
the $Xe$ (and $H_2$) condensate such impurities are absent . In the $O_2$
condensate size of the excimer molecules $(O_2)_2$ is larger than that of
the $(S)_2$ molecules and the superconducting transition temperature is
about $0.6K$ \cite{7}.

The HTSC materials may be considered as a some kind of the metal-dielectric
(MC) nanocomposites with complex type of bond and of structures, as a
``payment'' for their stability at normal conditions. The $X$ parameter of
the MC may be increased near the contact with metal (metal as a catalyst)
\cite{1.2}. It may simulate some effective pressure. The $O^{2-}$ oxygen ion
radius in compounds ($1.4\,\AA $) is close to that of the excited state
orbital $3s$ for the free atom ($1.6\,\AA $), and the critical $T_c$ may be
limited by the ``paramagnetic impurities''. A situation, which is
alternative to the case discussed in \cite{1,2} (electron pairs near the two
identical centers, for example $(Xe)_2$ , $(O_2)_2$), may be considered
also: electron pairs at monatomic centers, for example $O^{2-}$ ($3s^2$); $Ba
$ ($6s^2$). The transitions $(Ba^{2+};\,O^{2-})\leftrightarrow (Ba;O)$, like
in the intermediate valence systems may be taken into account too. It seems
to have some resemblance with the beginning of the Mott transition of $ns$
or $ns^2$ atoms (before the ordinary metal or dielectric appears, and the
bond energy of the electron ``pairs'' vanishes). The Bose condensation seems
to be less sensitive to disorder than the Fermi one.

A simple way to examine the discussed model is to carry out the joint
condensation of the $ns$ or $ns^2$ atoms together with the diamagnetic atoms
or molecules ($Xe$, $H_2$, $H_2O$, $NH_3$ and so on), and to obtain a
``physical'' model of the disordered MC. The ion implantation methods seems
to be applicable too.

At present time there are no strong evidence of the impossibility of the
SHTSC for the molecular condensates in spite of some inevitable magnetic
problems.

\end{document}